\documentclass[aps,pr,twocolumn,psfig,eqsecnum,epsf]{revtex4}
\usepackage{graphicx}
\usepackage{color}

\usepackage[version=3]{mhchem} 
\usepackage{amssymb}

\begin{document}
\title{\Large \bf Ion size effects on the electric double layer of a spherical particle in a realistic salt-free concentrated suspension}
\author{ \large Rafael Roa,$^1$ F\'elix Carrique,$^{1,}$\footnote{E-mail address: carrique@uma.es} and Emilio Ruiz-Reina$^2$\\} 
\affiliation{ $^1$ F\'{\i}sica Aplicada I, Universidad de M\'alaga, Spain,\\
$^2$ F\'{\i}sica Aplicada II, Universidad de M\'alaga, Spain.}
\date{\today}

\vspace{0.5cm} 
\begin{abstract}                
A new modified Poisson-Boltzmann equation accounting for the finite size of the ions valid for realistic salt-free concentrated suspensions has been derived, extending the formalism developed for pure salt-free suspensions [Roa \textit{et al.}, \textit{Phys. Chem. Chem. Phys.}, 2011,  \textbf{13}, 3960-3968] to real experimental conditions. These realistic suspensions include water dissociation ions and those generated by atmospheric carbon dioxide contamination, in addition to the added counterions released by the particles to the solution. The electric potential at the particle surface will be calculated for different ion sizes and compared with classical Poisson-Boltzmann predictions for point-like ions, as a function of particle charge and volume fraction. The realistic predictions turn out to be essential to achieve a closer picture of real salt-free suspensions, and even more important when ionic size effects are incorporated to the electric double layer description. We think that both corrections have to be taken into account when developing new realistic electrokinetic models, and surely will help in the comparison with experiments for low-salt or realistic salt-free systems.
\end{abstract} 
\maketitle
\vspace{0.5cm}


\section{Introduction}
Many efforts have been devoted in the past, and still continue nowadays, with the aim of improving our knowledge of the electric double layer (EDL) surrounding a charged particle in a colloidal suspension \cite{Dukhin1974,Lyklema1995,Hunter1995,Masliyah2006}. It is a well known fact that many non-equilibrium phenomena in this kind of systems are extremely sensitive to the specific properties of such EDL. For many years different electrokinetic models for colloidal suspensions have been derived. Most of them are based on the classical Poisson-Boltzmann equation (PB), which is a mean-field theory with a reasonable success in representing the ionic concentration profiles at low to moderately charged interfaces in electrolyte solutions. However, one of its main drawbacks concerns the absence of size for the ions, which, for highly charged particles, yields to unphysically high counterion concentration profiles near such interfaces. In addition, the PB treatment neglects ion-ion correlations, which simplifies the real scenario. Thus, the interaction on a particular ion is just represented by that in a mean-field, which is considered to be a poor description of real interactions taking place inside the EDL by the microscopic models, in many cases \cite{GonzalezTovar1989,IbarraArmenta2009}.

On the other hand, most of the theoretical studies with colloidal suspensions corresponds to the dilute regime in particle concentration, in spite of it is the concentrated one that deserves more interest due to its many industrial applications \cite{OBrien1990,Ohshima2006,Dukhin1999,Carrique2003}. The reason has to do with the larger complexity associated with the electrohydrodynamic particle-particle interactions in such concentrated systems, that are  very difficult to manage theoretically. This has encouraged us to consider the latter particle-particle interactions and the possibility of overlapping between adjacent double layers which will be unavoidably present with high particle concentrations. In many typical cases, the presence of an external salt added to the system gives rise to an effective screening effect on repulsive electrostatic particle-particle interactions, depending on the salt concentration, which are mainly responsible, for example, of the generation of colloidal crystals or glasses. Thus, it would be of worth to study systems with low screening regime for such interactions.

Those systems are named salt-free because of the absence of added external salt. The formation of colloidal crystals is easier in this kind of systems, even at sufficiently low particle volume fractions \cite{Sood1991,Medebach2003}. Of course, these salt-free systems contain ions in solution, the so-called ``added counterions" stemming from the particles as they get charged, that counterbalance their surface charge preserving the electroneutrality \cite{RuizReina2008,Carrique2006,Carrique2008,Carrique2008b,RuizReina2007}. These salt-free suspensions have acquired a renovated interest in the last few years due primarily to the special phenomenology they show related to the colloid crystals. 

The authors have recently developed an EDL model of a spherical particle in a salt-free suspension and a robust and efficient mathematical treatment to numerically solved the PB equation, which for that case may become integrodifferential due to the coupling between particle charge and ionic countercharge in the solution. In addition, realistic conditions associated to the presence of additional ions dissolved in the liquid medium, like those associated with water dissociation and possible atmospheric contamination, were included in the analysis \cite{RuizReina2008}. The aim was to improve the EDL model that could be used to develop nonequilibrium models in concentrated systems, starting from the salt-free ones because of their special theoretical interest. Thus, static and dynamic electrophoresis and complex conductivity and dielectric response models were developed by the authors according to such EDL representation \cite{Carrique2009,Carrique2009b,RuizReina2010,Carrique2010}. The first conclusion that can be drawn from these studies is the large magnitude of the realistic corrections for any of the latter properties. As a consequence, the neglecting of such corrections could lead to poor comparisons with experimental results. We can assure that the better the EDL representation of a real charged particle in a colloidal suspension, the better predictions the macroscopic nonequilibrium models will be able to do.

It is important to realize that the mean-field PB approximation neglects ion-ion correlations, which may be responsible of some phenomena like overcharging or charge inversion that historically have not been predicted within this classical framework \cite{Lyklema2009}. A first attempt to include some of these correlations concerns those linked to the ion excluded volume. The point-like ions of the PB approach are now considered as ions with finite size through a modified Poisson-Boltzmann equation (MPB) \cite{Bikerman1942,Outhwaite1983,Adamczyk1996,Borukhov1997,Borukhov2000,Borukhov2004,Lue1999,KraljIglic1996,Bohinc2001,Biesheuvel2007,Bazant2009,LopezGarcia2007,LopezGarcia2008,ArandaRascon2009,ArandaRascon2009b,Bhuiyan2004,Bhuiyan2009}. It has been recently shown that these MPB predictions of EDL equilibrium properties for dilute suspensions with a monovalent salt agree well with those obtained by Monte Carlo simulations or microscopic models, although the agreement worsens as the valence of the counterions increases. Unfortunately, this latter MPB model showed to be unable to predict the overcharging phenomena whereas Monte Carlo simulations or other microscopic models succeeded \cite{IbarraArmenta2009}.

Very recently, L\'opez-Garc\'ia \textit{et al.} \cite{LopezGarcia2010} have presented a mean-field MPB approach that includes finite ion size corrections with the additional assumption of a different distance of closest approach to the particle surface for counterions than for coions. They show that this model can predict charge inversion in the case of high electrolyte concentrations and counterion valence. We think that this is a very important result because it demonstrates that a mean-field theory can predict charge inversion, while microscopic models or Monte Carlo simulations achieve similar results by considering full ion-ion correlations.

In this paper we will use an analogous MPB approach as the one previously derived for the pure salt-free case that included ionic size effects \cite{Roa2011}. It generalizes that of Borukhov \cite{Borukhov2004} for dilute salt-free systems to the concentrated one, with the additional incorporation of an excluded region of a hydrated ion radius in contact with the particle. This latter issue has been shown by Aranda-Rasc\'on \textit{et al.} \cite{ArandaRascon2009} to provide a better representation of the solid-liquid interface and more reliable electrokinetic predictions. We will extend the finite ion size formalism to the case of realistic concentrated salt-free suspensions, already studied by the authors for point-like ions \cite{RuizReina2008}. As both corrections have considerably modified the standard PB predictions, it is mandatory to know which are the predictions of a general mean-field approach including both, realistic salt-free corrections and finite ion size corrections. This is a difficult task because of the increasing numerical problems arisen when simultaneously take them into account in the resulting MPB equations derived for each case. It is also important to point out that the numerical instabilities progressively grow because of the iterative methods, unlike previous models for realistic salt-free systems where commonly no iterative methods were necessary. In the following sections we will present the theoretical model for this kind of systems. The resulting equations will be numerically solved and the equilibrium potential at the particle surface will be analyzed upon changing particle volume fraction, particle surface charge density, and size of the ions. In order to show the realm of the finite ion size effect in realistic salt-free suspensions, the results will be compared with MPB predictions that do not take into account a finite distance of closest approach to the particle surface, and also with standard PB predictions for point-like ions.

\section{Theory}

\subsection{The cell model}
To account for the interactions between particles in concentrated suspensions, a cell model is used (bare Coulomb interactions among particles are included in an average sense, but ions-induced interactions between particles as well as ion-ion correlations, are ignored). For details about the cell model approach see the excellent review of Zholkovskij \textit{et al.} \cite{Zholkovskij2007}. This approach has been successfully used by the authors in the study of DC and AC electrokinetics and rheological properties of pure \cite{Carrique2006,Carrique2008,Carrique2008b,RuizReina2007} and realistic \cite{RuizReina2008,Carrique2009,Carrique2009b,RuizReina2010,Carrique2010} salt-free suspensions with point-like ions. We have learned from those works how important the description of the EDL is for the non-equilibrium theoretical responses. 

\begin{figure}[b]
\centering
  \includegraphics[height=4.5cm]{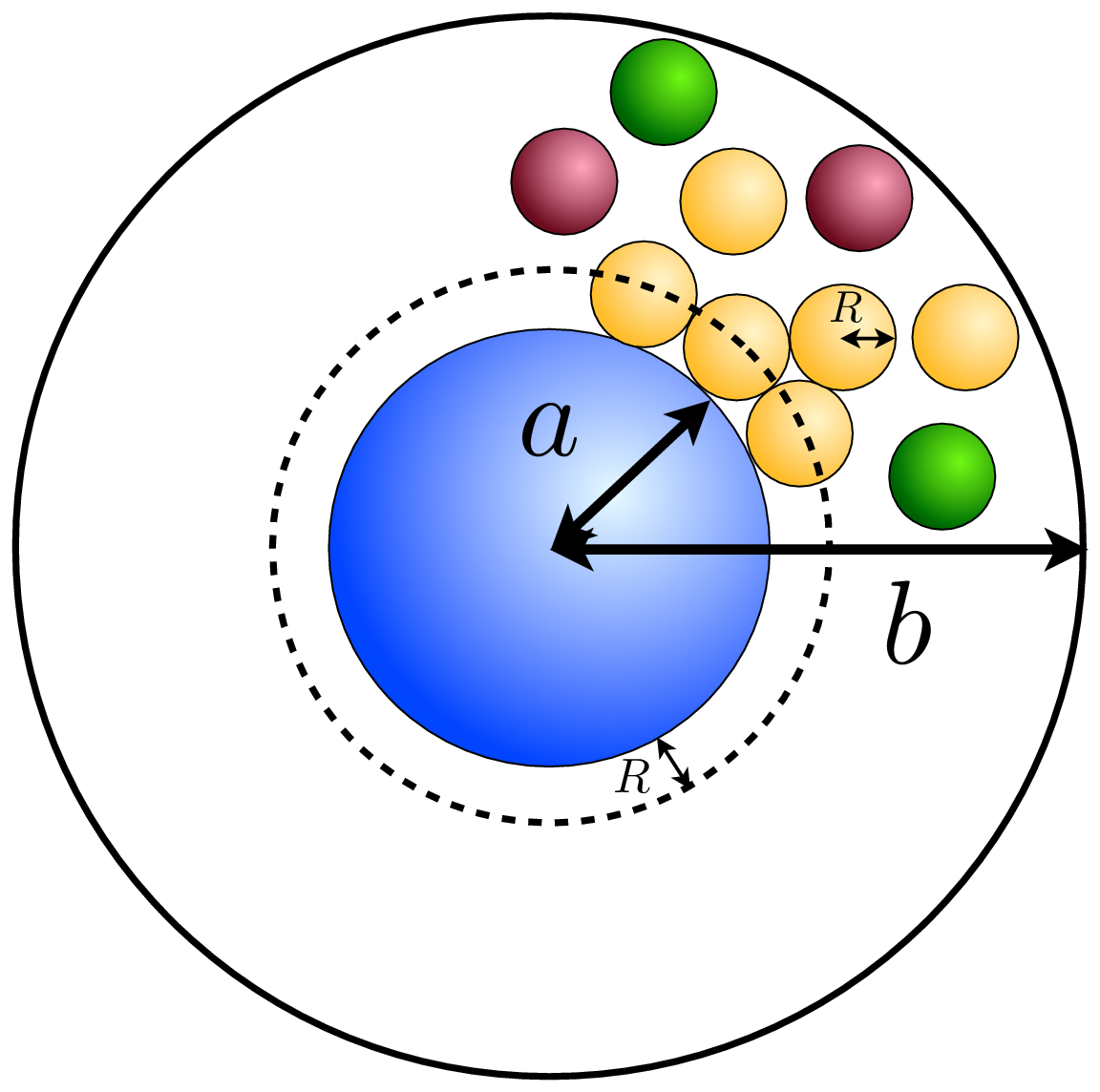}
  \caption{Cell model in a realistic salt-free suspension including a distance of closest approach of the ions to the particle surface.}
  \label{fgr:cellreal}
\end{figure}

Concerning the  cell model, Figure \ref{fgr:cellreal}, each spherical particle of radius $a$ is surrounded by a concentric shell of the liquid medium, having an outer radius $b$ such that the particle/cell volume ratio in the cell is equal to the particle volume fraction throughout the entire suspension, that is \cite{Happel1958}
\begin{equation}\label{phidef}
\phi=\left( \frac{a}{b}\right)^3
\end{equation}

The basic assumption of the cell model is that the macroscopic properties of a suspension can be obtained  from appropriate averages of local properties in a unique cell.

\subsection{Finite size of the ions}
In a very recent study we addressed the EDL of a pure salt-free suspension taking into account the finite size of the counterions \cite{Roa2011}. In this work we deal with realistic salt-free suspensions and, consequently, we will have more ionic species dissolved in the liquid medium, that will be coupled by appropriate chemical equilibrium mass-action equations. Our systems consist of aqueous suspensions deionized maximally without any electrolyte added during the preparation. Hence, in addition to the added counterions released by the particles to the solution, we must also consider the \ce{H+} and \ce{OH-} ions from water dissociation. Moreover, if the suspension is open to the atmosphere, there will be other ions produced by the atmospheric \ce{CO2} contamination. All these new ionic species could be coincident, or not, with that of the added counterions. Here, we will describe the general case of $N$ ionic species in the suspension. Details for each situation can be found in subsection \ref{subrealistic}. 

Let us consider a spherical charged particle of radius $a$ and surface charge density $\sigma$ immersed in a realistic salt-free medium with $N$ ionic species including that of the added counterions, that counterbalance its surface charge. 

In our description, the axes of the spherical coordinate system ($r$, $\theta$, $\varphi$) are fixed at the center of the particle. In the absence of any external field, the particle is surrounded by a spherically symmetrical charge distribution.

Within a mean-field approximation, the total free energy of the system, $F=U-TS$, can be written in terms of the equilibrium electric potential $\Psi(\mathbf{r})$ and the ionic concentration $n_i(\mathbf{r})$ of the different ionic species, $i=1, \ldots, N$, of the suspension. The configurational internal energy contribution $U$ is
\begin{multline}
  U=\int d\textbf{r}\Bigg{[}-\frac{\epsilon_0\epsilon_r}{2}|\nabla\Psi(\mathbf{r})|^2 \\
 +\sum_{i=1}^Nz_ien_i(\mathbf{r})\Psi(\mathbf{r})-\sum_{i=1}^N\mu_in_i(\mathbf{r})\Bigg{]}
\end{multline}

The first term is the self-energy of the electric field, where $\epsilon_0$ is the vacuum permittivity, and $\epsilon_r$ is the relative permittivity of the suspending medium. The next term is the sum of the electrostatic energies of the different ionic species in the electrostatic mean field, and the last term couples the system to a bulk reservoir, where $\mu_i$ is the chemical potential of the ionic species $i$.

The entropic contribution $-TS$ is
\begin{multline}
  -TS=k_BTn^{max}\int d\textbf{r}\Bigg{[}\sum_{i=1}^N\frac{n_i(\mathbf{r})}{n^{max}}\ln \left(\frac{n_i(\mathbf{r})}{n^{max}}\right) \\
  +\left(1-\sum_{i=1}^N\frac{n_i(\mathbf{r})}{n^{max}}\right)\ln \left(1-\sum_{i=1}^N\frac{n_i(\mathbf{r})}{n^{max}}\right)\Bigg{]}
\end{multline}
where $k_B$ is Boltzmann's constant, $T$  is the absolute temperature, and $n^{max}$ is the maximum possible ionic concentration due to the excluded volume effect, defined as $n^{max}=V^{-1}$, where $V$ is the average volume occupied by an ion in the solution. For simplicity, we assume that all types of ions have the same size, and therefore $n^{max}$ will take the same value for all of them. The first term inside the integral is the sum of the entropy of the different ionic species, and the second one is the entropy of the solvent molecules. This last term accounts for the ion size effect that modifies the classical Poisson-Boltzmann equation and was proposed earlier by Borukhov \textit{et al.} \cite{Borukhov1997}

The variation of the free energy $F=U-TS$ with respect to $\Psi(\mathbf{r})$ provides the Poisson equation
\begin{equation}\label{poisson}
\nabla^2\Psi(\mathbf{r})=-\frac{e}{\epsilon_0\epsilon_r}\sum_{i=1}^Nz_in_i(\mathbf{r})
\end{equation}
and the ionic concentration of the ionic species $i$ is obtained by performing the variation of the free energy with respect to $n_i(\mathbf{r})$, yielding 
\begin{equation}\label{ni}
n_i(\mathbf{r})=\frac{b_i\exp\left(-\frac{z_ie\Psi(\mathbf{r})}{k_BT}\right)}{\displaystyle1+\sum_{j=1}^N\frac{b_j}{n^{max}}\left[\exp\left(-\frac{z_je\Psi(\mathbf{r})}{k_BT}\right)-1\right]}
\end{equation}
where $b_i$ is an unknown coefficient that represents the ionic concentration of the species $i$ where the electric potential is zero.

Applying the spherical symmetry of the problem and combining Equations \ref{poisson} and \ref{ni}, we obtain the modified Poisson-Boltzmann equation (MPB) for the equilibrium electric potential
\begin{multline}\label{mpbreal}
\frac{\mathrm{d}^2\Psi(r)}{\mathrm{d}r^2}+\frac{2}{r}\frac{\mathrm{d}\Psi(r)}{\mathrm{d}r} \\
=-\frac{e}{\epsilon_0\epsilon_r}\frac{\displaystyle\sum_{i=1}^Nz_ib_i\exp\left(-\frac{z_ie\Psi(\mathbf{r})}{k_BT}\right)}{\displaystyle1+\sum_{i=1}^N\frac{b_i}{n^{max}}\left[\exp\left(-\frac{z_ie\Psi(\mathbf{r})}{k_BT}\right)-1\right]}
\end{multline}

We need two boundary conditions to solve the MPB equation. The first one is
\begin{equation}\label{dpot-b}
\frac{\mathrm{d}\Psi(r)}{\mathrm{d}r}\bigg |_{r=b}=0
\end{equation}
which derives from the electroneutrality condition of the cell and the application of Gauss theorem to the outer surface of the cell. The second one is
\begin{equation}\label{pot-b}
\Psi(b)=0
\end{equation}
that fixes the origin of the electric potential at $r=b$.

The MPB problem, Equations \ref{mpbreal}, \ref{dpot-b}, and \ref{pot-b}, can be solved iteratively using the electroneutrality condition of the cell and appropriate chemical reactions to find the unknown $b_i$ coefficients (see subsection \ref{subrealistic} for details on $b_i$ calculation). This kind of problem can be solved in a better way by using dimensionless variables \cite{RuizReina2008}, which are defined as
\begin{align}\label{adim}
x=\frac{r}{a}; \quad 
\tilde{\Psi}(x)=&\frac{e\Psi(r)}{k_BT}; \quad 
\tilde{\sigma}=\frac{ea}{\epsilon_0\epsilon_rk_BT}\sigma \nonumber\\
\tilde{b}_i=\frac{e^2a^2}{\epsilon_0\epsilon_rk_BT}b_i; & \quad 
\tilde{n}^{max}=\frac{e^2a^2}{\epsilon_0\epsilon_rk_BT}n^{max}
\end{align}
rewriting Equation \ref{mpbreal} as
\begin{multline}\label{mpbrealadim}
g(x)\equiv\frac{\mathrm{d}^2\tilde{\Psi}(x)}{\mathrm{d}x^2}+\frac{2}{x}\frac{\mathrm{d}\tilde{\Psi}(x)}{\mathrm{d}x} \\
=\frac{-\displaystyle\sum_{i=1}^Nz_i\tilde{b}_i\mathrm{e}^{-z_i\tilde{\Psi}(x)}}{\displaystyle1+\sum_{i=1}^N\frac{\tilde{b}_i}{\tilde{n}^{max}}\left(\mathrm{e}^{-z_i\tilde{\Psi}(x)}-1\right)}
\end{multline}
where we have defined the function $g(x)$. If we differentiate it, after a little algebra, we find that
\begin{multline}\label{gprimereal}
g'(x)+\frac{\displaystyle\sum_{i=1}^Nz_i^2\tilde{b}_i\mathrm{e}^{-z_i\tilde{\Psi}(x)}}{\displaystyle\sum_{i=1}^Nz_i\tilde{b}_i\mathrm{e}^{-z_i\tilde{\Psi}(x)}}g(x)\tilde{\Psi}'(x) \\
+\frac{1}{\tilde{n}^{max}}g^2(x)\tilde{\Psi}'(x)=0
\end{multline}
where the prime stands for differentiation with respect to $x$. In terms of the electric potential, Equation \ref{gprimereal} is rewritten as
\begin{multline}\label{mpb3real}
\tilde{\Psi}'''(x)+\frac{2}{x}\tilde{\Psi}''(x)-\frac{2}{x^2}\tilde{\Psi}'(x)+\tilde{\Psi}'(x)\left(\tilde{\Psi}''(x)+\frac{2}{x}\tilde{\Psi}'(x)\right)\\ 
\cdot\left[\frac{\displaystyle\sum_{i=1}^Nz_i^2\tilde{b}_i\mathrm{e}^{-z_i\tilde{\Psi}(x)}}{\displaystyle\sum_{i=1}^Nz_i\tilde{b}_i\mathrm{e}^{-z_i\tilde{\Psi}(x)}}+\frac{1}{\tilde{n}^{max}} \left(\tilde{\Psi}''(x)+\frac{2}{x}\tilde{\Psi}'(x)\right)\right]=0
\end{multline}
Equation \ref{mpb3real} is a nonlinear third-order differential equation that needs three boundary conditions to completely specify the solution. Two of them are provided by Equations \ref{dpot-b} and \ref{pot-b}, which now read
\begin{equation}\label{bcmpb31}
\tilde{\Psi}'(h)=0; \quad \tilde{\Psi}(h)=0 
\end{equation}
where $h=(b/a)=\phi^{-1/3}$ is the dimensionless outer radius of the cell. The third one specifies the electrical state of the particle, and can be obtained by applying Gauss theorem to the outer side of the particle surface $r=a$ 
\begin{equation}\label{dpot-a}
\frac{\mathrm{d}\Psi(r)}{\mathrm{d}r}\bigg |_{r=a}=-\frac{\sigma}{\epsilon_0\epsilon_r}
\end{equation}

Its dimensionless form is
\begin{equation}\label{bcmpb32}
 \tilde{\Psi}'(1)=-\tilde{\sigma}
\end{equation}

Many theoretical and experimental studies have recently explored the use of either constant surface potential or constant surface charge boundary conditions. These conditions, although making the mathematical treatment simpler, represent only limiting or idealized cases. Many biological and artificial particles have their surface charge associated with some degree of dissociation of functional groups which depend on the nearby environment. Constant surface potential and constant surface charge models would correspond, respectively, to the cases when the dissociation reactions of the functional groups are infinitely fast and infinitely slow \cite{Krozel1992}. Also, some authors have derived an hybrid surface charge model to account for the electrical state of the particles \cite{Lee2002}, which is a generalization of the conventional constant surface potential and constant surface-charged density models. Our model can be modified to include charge regulation mechanisms at the particle surface. 

It is very common in the literature to use the surface charge or the surface potential as a boundary condition at the particle surface when solving the equilibrium Poisson-Boltzmann equation, and both of them are valid. When it comes to concentrated suspensions, we prefer to use the particle surface charge as a boundary condition because in many cases of interest the particle charge is a property that can be determined experimentally. Moreover, the use of commercially available latex suspensions with fully dissociated surface electrical groups, have led us to choose the constant surface charge boundary condition. In the near future our intention is to make electrokinetic predictions based on our model that can be checked against experimental data, and the latex suspensions are probably the most promising ones for that task. 

Besides, the surface potential depends on the choice of the potential origin, and in the case of concentrated suspensions there is not a standard criterium for this choice. As stated in Equation \ref{pot-b} we have chosen it at the outer surface of the cell, $r=b$.

If we consider point-like ions, $n^{max}=\infty$, Equation \ref{mpb3real} generates the expressions obtained by Ruiz-Reina and Carrique \cite{RuizReina2008} for realistic salt-free suspensions. Also, if we evaluate Equation \ref{mpb3real} considering that we have just one ionic species, the added counterions that counterbalance the particle surface charge, the equation reduces to the one obtained by the authors \cite{Roa2011} for pure salt-free suspensions including finite ion size effects. 

It can be easily demonstrated that solving the third-order problem is mathematically equivalent to finding the solution of the MPB problem, that is, any function satisfying Equations \ref{mpb3real}, \ref{bcmpb31} and \ref{bcmpb32} also satisfies Equations \ref{mpbreal}, \ref{dpot-b}, and \ref{pot-b}, and vice versa. In both cases, we will use an appropriate iterative process for the calculation of the $b_i$ coefficients in the resolution of the differential equation. We set initial values for the different $b_i$ coefficients, but the use of the third order problem with the addition of the boundary condition Equation \ref{bcmpb32}, improves the convergency giving rise to a better numerical resolution. 

Equations \ref{mpb3real}, \ref{bcmpb31} and \ref{bcmpb32} form a boundary value problem that can be numerically solved by using the MATLAB routine bvp4c \cite{Kierzenka2001}, that computes the solution with a finite difference method by the three-stage Lobatto IIIA formula. This is a collocation method that provides a $C^1$ solution that is fourth order uniformly accurate at all the mesh points. The resulting mesh is non-uniformly spaced out and has been chosen to fulfill the admitted error tolerance (taken as $10^{-6}$ for all the calculations).

Once we have found the electric potential $\tilde{\Psi}(x)$ with the iterative process for the calculation of the $b_i$ coefficients, we can obtain the ionic concentration $n_i(r)$ for the different ionic species (see subsection \ref{subrealistic} for details).

\subsection{Excluded region in contact with the particle}
Following the work of Aranda-Rasc\'on \textit{et al.} \cite{ArandaRascon2009}, we incorporate a distance of closest approach of the ions to the particle surface, resulting from their finite size. As we said before, for the sake of simplicity we have considered that all ions have the same size. We assume that ions cannot come closer to the surface of the particle than the chosen effective hydration ionic radius, $R$, and, therefore, the ionic concentration will be zero in the region between the particle surface, $r=a$, and the spherical surface, $r=a+R$, defined by the ionic effective radius.

The whole electric potential $\Psi(r)$ is now determined by combining Laplace's and MPB equations into the following MPBL stepwise equation
\begin{equation}\label{laplace2real}
\begin{cases}
\frac{\mathrm{d}^2\Psi(r)}{\mathrm{d}r^2}+\frac{2}{r}\frac{\mathrm{d}\Psi(r)}{\mathrm{d}r}=0 & \text{ \ } a\leq r \leq a+R \\
& \\
\mathrm{Equation}\ \ref{mpbreal} & \text{ \ } a+R\leq r \leq b
\end{cases}
\end{equation}

We must impose the continuity of the potential and of its first derivative at the surface $r=a+R$, in addition to boundary conditions, Equations \ref{dpot-b} and \ref{pot-b}. The continuity of the first derivative comes from the continuity of the normal component of the electric displacement at that surface. Thus, in the region in contact with the particle $[a,\ a+R]$, we are solving Laplace's equation, and, in the region $[a+R,\ b]$, the MPB equation that we obtained previously in Equation \ref{mpbreal}.

As we have seen before, changing the system of second order differential equations, Equation \ref{laplace2real}, into one of third order, hugely simplify the resolution process
\begin{equation}\label{laplace3real}
\begin{cases}
\tilde{\Psi}'''(x)+\frac{2}{x}\tilde{\Psi}''(x)-\frac{2}{x^2}\tilde{\Psi}'(x)=0 & \text{ \ } 1\leq x \leq 1+\delta \\
& \\
\mathrm{Equation}\ \ref{mpb3real} & \text{ \ } 1+\delta \leq x \leq h
\end{cases}
\end{equation}
where $\delta=R/a$ and we have use dimensionless variables. The boundary conditions needed to completely close the problem are
\begin{align}\label{bclaplace3}
\tilde{\Psi}'_L(1)=-\tilde{\sigma} \quad & \quad \tilde{\Psi}'_L(1+\delta)=\frac{-\tilde{\sigma}}{(1+\delta)^2} \nonumber\\ 
\tilde{\Psi}_P(h)=0 \quad & \quad \tilde{\Psi}'_P(h)=0 \nonumber \\
\tilde{\Psi}_L(1+\delta)=\tilde{\Psi}_P(1+\delta) \quad & \quad \tilde{\Psi}'_L(1+\delta)=\tilde{\Psi}'_P(1+\delta)
\end{align}
where subscript $L$ refers to the region in which the potential is calculated using Laplace's equation, and subscript $P$ refers to the region in which we evaluate the MPB equation.

The spherical solution of Laplace equation in the region free of charge is
\begin{equation}
 \tilde{\Psi}_L(x)=\frac{\tilde{\sigma}}{x}+K
\end{equation}
where $K$ is a constant. The potential difference in the Laplace layer becomes
\begin{equation}\label{shift}
 \Delta\tilde{\Psi}_L=\tilde{\Psi}_L(1)-\tilde{\Psi}_L(1+\delta)=\frac{\tilde{\sigma}\delta}{1+\delta}\approx\tilde{\sigma}\delta
\end{equation}
where we have assumed that $a\gg R$. We solve now the MPB problem only in the $[1+\delta,h]$ region, instead of the $[1,h]$ range that is used in the absence of a Laplace layer. Both solutions are very similar because $\delta\ll h$ and $\delta\ll 1$ and, consequently, we have 
\begin{equation}
 \tilde{\Psi}_P(1+\delta)\approx\tilde{\Psi}(1)
\end{equation}
where $\tilde{\Psi}(1)$ is the electric potential at the particle surface in the MPB case without any excluded region in contact with the particle. As a result, we expect that the surface potential will be shifted in accordance  with Equation \ref{shift} when we introduce the Laplace layer, in comparison with the MPB problem. The shift depends approximately linearly on the Laplace region thickness $\delta$ and, therefore, it remains roughly constant upon changing the volume fraction.

The complete electric potential $\tilde{\Psi}(x)$ is obtained numerically using Equations \ref{laplace3real} and \ref{bclaplace3}, including an iterative process for the calculation of the $b_i$ coefficients. Once the electric potential is found, the ionic concentrations $n_i(r)$ for the different ionic species can be derived (see subsection \ref{subrealistic} for details).

\subsection{Particularization of the MPBL equation for realistic salt-free concentrated suspensions}\label{subrealistic}

In a previous work \cite{Roa2011}, the authors considered the case of pure salt-free concentrated suspensions including ion size effects. In this subsection, we will focus our treatment to realistic salt-free concentrated suspensions, with the consideration of ions from water dissociation and atmospheric contamination, in addition to the added counterions. This will be carried out by using the generalized MPBL stepwise equation, Equation \ref{laplace2real}, or in a better way, its third order equivalent version given by Equation \ref{laplace3real}, previously obtained.

\subsubsection{Added counterions and water dissociation}

Let us consider that, in addition to the added counterions stemming from the particle charging process, there are also \ce{H+} and \ce{OH-} ions coming from water dissociation in the liquid medium. This will always occur in aqueous suspensions. The equilibrium mass-action equation for water dissociation, which we assume to hold at the outer surface of the cell, is
\begin{equation}\label{bHbOH}
[\ce{H+}][\ce{OH-}]=K_\mathrm{w} \quad \Rightarrow \quad \tilde{b}_\ce{H+}\tilde{b}_\ce{OH-}=\tilde{K}_\mathrm{w}
\end{equation}
where the square brackets stand for the molar concentration, $K_\mathrm{w}=10^{-14}$ mol$^2$/L$^2$ is the water dissociation constant at room temperature, 298.15 K, and $\tilde{K}_\mathrm{w}$ is a dimensionless quantity defined by
\begin{equation}
\tilde{K}_\mathrm{w}=\left(\frac{10^3N_Ae^2a^2}{\epsilon_0\epsilon_rk_BT}\right)^2K_\mathrm{w}
\end{equation}
with $N_A$ the Avogadro constant.

We can distinguish between two cases, (a) when the added counterions are \ce{H+} or \ce{OH-} ions, and (b) when they are of a different ionic species. The distinction is important because in the (a) case, the added counterions will enter in the equilibrium reaction equation for water dissociation, whereas in the (b) case, they do not.

\paragraph*{\textit{Case a.}} In this case we have two different ionic species. Evaluating Equation \ref{mpbrealadim}, particularized just for \ce{H+} and \ce{OH-} ions, at $x=h$ we obtain
\begin{equation}\label{bHbOH-2}
\tilde{\Psi}''(h)=-z_\ce{H+}\tilde{b}_\ce{H+}-z_\ce{OH-}\tilde{b}_\ce{OH-}
\end{equation}

Using the relation between both $\tilde{b}_\ce{H+}$ and $\tilde{b}_\ce{OH-}$ coefficients given by Equation \ref{bHbOH}, we can write Equations \ref{laplace3real} and \ref{bHbOH-2} in terms of just one unknown coefficient, $\tilde{b}_\ce{H+}$. The iterative process needed for the numerical resolution of the third order version of the MPB problem remains as follows. We choose an initial guess for the $\tilde{b}_\ce{H+}$ coefficient, say $\tilde{b}_\ce{H+}^{(0)}=0$ or, in a more accurate way, the value obtained for point-like ions, and solve Equation \ref{laplace3real} with the boundary conditions given by Equation \ref{bclaplace3}. We obtain the solution $\tilde{\Psi}^{(0)}(x)$, and then, we use Equations \ref{bHbOH} and \ref{bHbOH-2} to find a new value $\tilde{b}_\ce{H+}^{(1)}$, which will give us $\tilde{\Psi}^{(1)}(x)$ using Equations \ref{laplace3real} and \ref{bclaplace3} again. The numerical iterative process is repeated until the relative variation of the electric potential at the particle surface is lower than a prescribed quantity. Although an iterative process has been used to obtain the solution to Equation \ref{laplace3real} with boundary conditions, Equation \ref{bclaplace3}, and Equations \ref{bHbOH} and \ref{bHbOH-2}, this procedure is much better than the original and equivalent iterative problem defined by Equation \ref{laplace2real}. The improved convergency and superior numerical efficiency that are obtained when computing the third order problem lie in the facts that all of the intermediate solutions $\tilde{\Psi}^{(n)}(x)$ of the iterative method have the correct slope $\tilde{\Psi}'(1)=-\tilde{\sigma}$ at the particle surface. This is not true if we use the original scheme because in that case the slope at the particle surface is not determined by any condition.

\paragraph*{\textit{Case b.}} In this case we have three different ionic species. Evaluating Equation \ref{mpbrealadim}, particularized for the added counterions, of valence $z_c$, and the ions \ce{H+} and \ce{OH-}, at $x=h$ we obtain
\begin{equation}\label{bcbHbOH}
\tilde{\Psi}''(h)=-z_c\tilde{b}_c-z_\ce{H+}\tilde{b}_\ce{H+}-z_\ce{OH-}\tilde{b}_\ce{OH-}
\end{equation}

The added counterions counterbalance the overall charge on the particle surface
\begin{equation}\label{wden}
\sigma=-\int_1^h\frac{z_cb_c\mathrm{e}^{-z_c\Psi(x)}}{1+\displaystyle\sum_{i=1}^N\frac{\tilde{b}_i}{\tilde{n}^{max}}\left(\mathrm{e}^{-z_ie\Psi(x)}-1\right)}x^2\mathrm{d}x
\end{equation}
whereas the number of \ce{H+} and \ce{OH-} must be equal due to the electroneutrality of the cell. 

Using the relation between both $\tilde{b}_\ce{H+}$ and $\tilde{b}_\ce{OH-}$ coefficients given by Equation \ref{bHbOH}, we can write Equations \ref{laplace3real}, \ref{bcbHbOH}, and \ref{wden} in terms of two unknown coefficients, $\tilde{b}_c$ and $\tilde{b}_\ce{H+}$. The iterative process to obtain the solution to Equation \ref{laplace3real} with boundary conditions, Equation \ref{bclaplace3}, and Equations \ref{bHbOH}, \ref{bcbHbOH}, and \ref{wden}, is similar to the one described before, but in this case we have the two unknown coefficients, $\tilde{b}_c$ and $\tilde{b}_\ce{H+}$, to be determined iteratively. 

\subsubsection{Added counterions, water dissociation, and atmospheric contamination}
Let us consider now that, in addition to the added counterions and the \ce{H+} and \ce{OH-} ions coming from water dissociation, there are also present ions stemming from the atmospheric \ce{CO2} contamination in the liquid medium. This will always occur in aqueous suspensions in contact with the atmosphere; the \ce{CO2} gas diffused into the suspension combines with water molecules to form carbonic acid \ce{H2CO3}, and then, the following dissociation reactions take place
\begin{align}
\ce{H2CO3} & \rightleftarrows \ce{H+} + \ce{HCO3-}  \\
\label{CO32-}
\ce{HCO3-} &\rightleftarrows  \ce{H+} + \ce{CO3^=}
\end{align}
with equilibrium dissociation constants $K_1=4$.$47\cdot10^{-7}$ mol/L and $K_2=4$.$67\cdot10^{-11}$ mol/L at room temperature, 298.15 K, respectively. The concentration of \ce{H2CO3} molecules in water can be calculated from the solubility and the partial pressure of \ce{CO2} in standard air. For a temperature of 298.15 K and an atmospheric pressure of 101300 Pa, the concentration of carbonic acid is approximately $[\ce{H2CO3}]=1$.$08\cdot10^{-5}$ mol/L, being its particular value dependent on  the local environmental conditions. The dimensionless dissociation constants and \ce{H2CO3} concentration are
\begin{align}
\tilde{K}_1=\frac{10^3N_Ae^2a^2}{\epsilon_0\epsilon_rk_BT}K_1; & \quad 
\tilde{K}_2=\frac{10^3N_Ae^2a^2}{\epsilon_0\epsilon_rk_BT}K_2 \nonumber\\
\tilde{N}_\ce{H2CO3}=&\frac{10^3N_Ae^2a^2}{\epsilon_0\epsilon_rk_BT}N_\ce{H2CO3}
\end{align}
where all the values are taken in S.I. units. The equilibrium mass-action equations, which we assume to hold at the outer surface of the cell, are
\begin{align}\label{bHbOHbHCO3}
\frac{[\ce{H+}][\ce{HCO3-}]}{[\ce{H2CO3}]}=K_1 & \quad \Rightarrow \quad \frac{\tilde{b}_\ce{H+}\tilde{b}_\ce{HCO3-}}{\tilde{N}_\ce{H2CO3}}=\tilde{K}_1  \\
\frac{[\ce{H+}][\ce{CO3^=}]}{[\ce{HCO3-}]}=K_2 & \quad \Rightarrow \quad \frac{\tilde{b}_\ce{H+}\tilde{b}_\ce{CO3^=}}{\tilde{b}_\ce{HCO3-}}=\tilde{K}_2
\end{align} 

Hereafter, the second dissociation reaction, Equation \ref{CO32-}, will be neglected because the terms associated to the ion $\ce{CO3^=}$ that appears in Equation \ref{laplace3real} are several orders of magnitude lower than those due to the ion $\ce{HCO3-}$, in accordance with what the authors showed in a previous work \cite{RuizReina2008}.

Once more, we can distinguish between two cases, (a) when the added counterions are coincident with one of the ionic species in the system (\ce{H+}, \ce{OH-} or \ce{HCO3-}) and (b) when they are of a different ionic species. In the (a) case, the added counterions will enter in one of the equilibrium dissociation equations, whereas in the (b) case, they do not.

\paragraph*{\textit{Case a.}} In this case we have three different ionic species. Evaluating Equation \ref{mpbrealadim}, particularized for \ce{H+}, \ce{OH-} and \ce{HCO3-} ions, at $x=h$ we obtain
\begin{equation}\label{bHbOHbHCO3-2}
\tilde{\Psi}''(h)=-z_\ce{H+}\tilde{b}_\ce{H+}-z_\ce{OH-}\tilde{b}_\ce{OH-}-z_\ce{HCO3-}\tilde{b}_\ce{HCO3-}
\end{equation}

Using the relations between coefficients $\tilde{b}_\ce{H+}$, $\tilde{b}_\ce{OH-}$ and $\tilde{b}_\ce{HCO3-}$ given by Equations \ref{bHbOH} and \ref{bHbOHbHCO3}, we can write Equations \ref{laplace3real} and \ref{bHbOHbHCO3-2} in terms of just one unknown coefficient, $\tilde{b}_\ce{H+}$. The iterative process to obtain the solution to Equation \ref{laplace3real} with boundary conditions, Equation \ref{bclaplace3}, and Equations \ref{bHbOH}, \ref{bHbOHbHCO3}, and \ref{bHbOHbHCO3-2}, is similar to the case (a) for added counterions and water dissociation described before.  

\paragraph*{\textit{Case b.}} In this case we have four different ionic species. Evaluating Equation \ref{mpbrealadim}, particularized for the added counterions, of valence $z_c$, and the ions \ce{H+}, \ce{OH-} and \ce{HCO3-}, at $x=h$ we obtain
\begin{equation}\label{bcbHbOHbHCO3}
\tilde{\Psi}''(h)=-z_c\tilde{b}_c-z_\ce{H+}\tilde{b}_\ce{H+}-z_\ce{OH-}\tilde{b}_\ce{OH-}-z_\ce{HCO3-}\tilde{b}_\ce{HCO3-}
\end{equation}

The added counterions counterbalance the overall charge on the particle surface, as in Equation \ref{wden}, whereas the number of ions \ce{H+}, \ce{OH-} and \ce{HCO3-} must balance due to the electroneutrality of the cell. 

Using the relations between coefficients $\tilde{b}_\ce{H+}$, $\tilde{b}_\ce{OH-}$ and $\tilde{b}_\ce{HCO3-}$ given by Equations \ref{bHbOH} and \ref{bHbOHbHCO3}, we can write Equations \ref{laplace3real}, \ref{bcbHbOHbHCO3}, and \ref{wden} in terms of two unknown coefficients, $\tilde{b}_c$ and $\tilde{b}_\ce{H+}$. The iterative process to obtain the solution to Equation \ref{laplace3real} with boundary conditions, Equation \ref{bclaplace3}, and Equations \ref{bHbOH}, \ref{bHbOHbHCO3}, \ref{bcbHbOHbHCO3}, and \ref{wden}, is similar to the case (b) for added counterions and water dissociation described before.

\section{Results and discussion}

For all the calculations, the temperature $T$ has been taken equal to 298.15 K and the relative electric permittivity of the suspending liquid $\epsilon_r=$ 78.55, which coincides with that of the deionised water. Also, the valence of the added counterions $z_c$ has been chosen equal to $+1$, when they are of a different ionic species as that of the ions stemming from water dissociation or atmospheric \ce{CO2} contamination, and the particle radius $a=100$ nm. Other values for $z_c$ could have been chosen. The model for point-like ions is able to work with any value of $z_c$ in the Poisson-Boltzmann equation, but we think that the predictions of this model will be less accurate in the case of multivalent counterions, since it is based on a mean-field approach that does not consider ion-ion correlations, which are increasingly more important as the ion valence grows. Nevertheless, when we take into account the finite size of the ions, the main objective of this work, we include correlations associated with the ionic excluded volume, solving partially this problem because we are still not considering electrostatic ion-ion correlations. 

For the sake of simplicity, we assume that  the average volume occupied by an ion is $V=(2R)^{3}$, being $2R$ the chosen effective ionic diameter. With this consideration, the maximum possible ionic concentration due to the excluded volume effect is $n^{max}=(2R)^{-3}$. This corresponds to a cubic package (52\% packing). In molar concentrations, the values used in the calculations, $n^{max}=$ 22, 4 and 1.7 M, correspond approximately to effective ionic diameters of $2R=$ 0.425, 0.75 and 1 nm, respectively. These are typical hydrated ionic diameters \cite{Israelachvili1992}.

We will discuss the results obtained from three different models, the classical Poisson-Boltzmann equation (PB), the modified Poisson-Boltzmann equation by the finite ion size effect (MPB), and the MPB equation including also a distance of closest approach of the ions to the surface of the charged particle (MPBL). Besides, we will consider different realistic salt-free concentrated suspensions: suspensions with added counterions coincident, WDC (read it as Water Dissociation Coincident), or not, WDNC (Water Dissociation Non-Coincident), with the ions that stem from water dissociation, \textit{i.e.}, \ce{H+} or \ce{OH-}; and suspensions with added counterions coincident, WDACC (read it as Water Dissociation Atmospheric Contamination Coincident), or not, WDACNC (Water Dissociation Atmospheric Contamination Non-Coincident), with the ions that stem from water dissociation or atmospheric contamination, \textit{i.e.}, \ce{H+}, \ce{OH-}, \ce{HCO3-}. We will compare them with the results of pure salt-free suspensions with only added counterions (SF), obtained by the authors in a previous work \cite{Roa2011}.

\begin{figure}[t!]
\centering
  \includegraphics[height= 6.4cm]{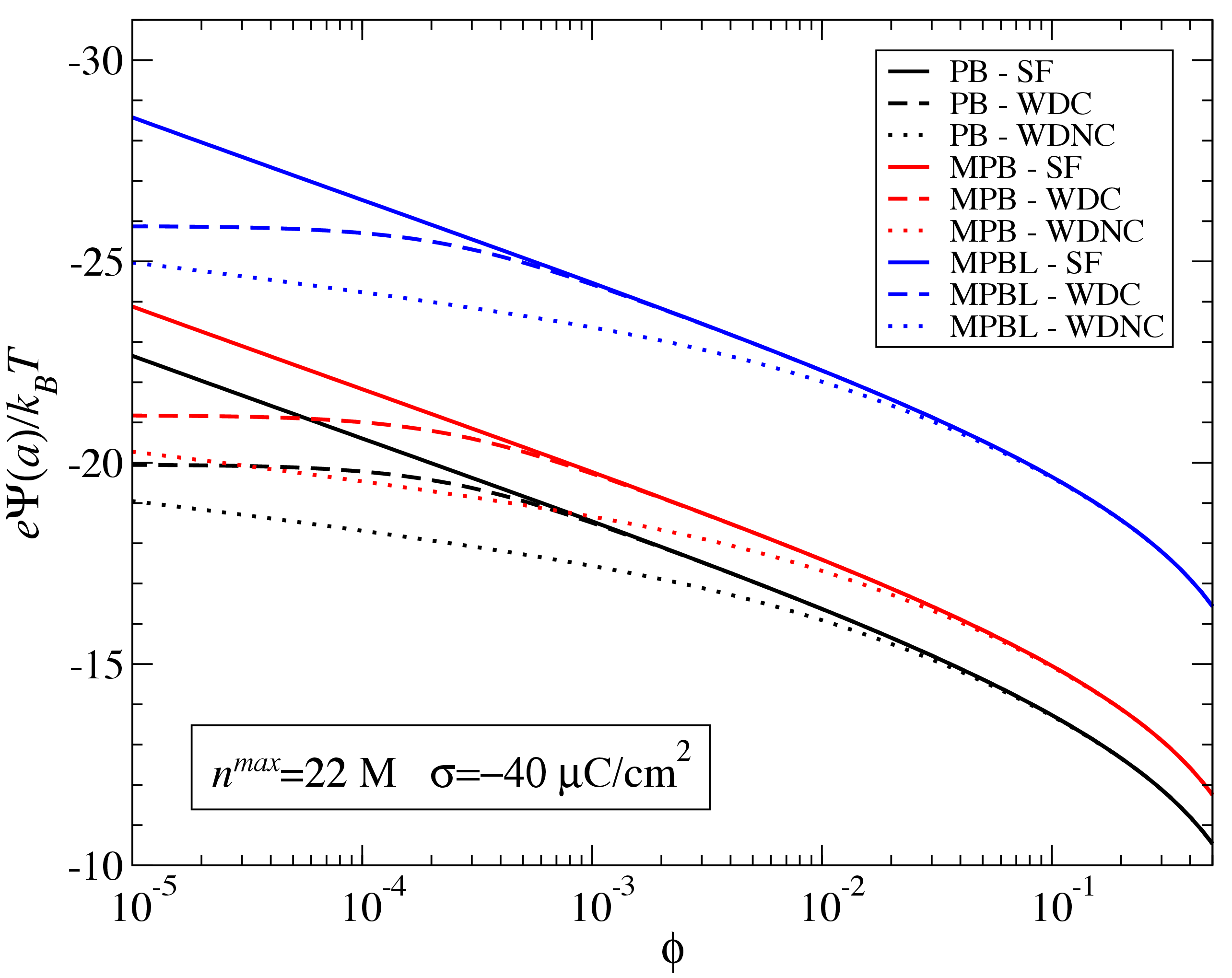}
  \caption{Dimensionless surface electric potential as a function of particle volume fraction, considering (blue lines, MPBL) or not (red lines, MPB) the excluded region in contact with the particle. Black lines show the results of the PB equation. Solid lines: only added counterions \ce{H+} (SF). Dashed lines: added counterions \ce{H+} with water dissociation ions (WDC). Dotted lines: added counterions different than \ce{H+} with $z_c=1$ and water dissociation ions (WDNC).}
  \label{fgr:WDC_WDNC_PHI}
\end{figure}

Fig. \ref{fgr:WDC_WDNC_PHI} shows the dimensionless equilibrium electric potential at the surface of the particle for a wide range of particle volume fractions. We display in black, red and blue lines the predictions of the PB, MPB and MPBL equations, respectively. Solid lines account for the results of pure salt-free suspensions with only added counterions \ce{H+} (SF).  Dashed lines stand for realistic WDC suspensions with \ce{H+} ions as added counterions, and dotted lines for realistic WDNC suspensions with added counterions different than \ce{H+} with $z_c=1$. The particle surface charge density have been chosen equal to $-$40 $\mu$C/cm$^2$, and the maximum possible ionic concentration due to the excluded volume effect $n^{max}=22$ M.

We can observe in Fig. \ref{fgr:WDC_WDNC_PHI} that for high particle concentration, the results for realistic suspensions with water dissociation ions are equal to those for pure salt-free suspensions, because the added counterions completely mask the influence of water dissociation ions, due to its larger concentration. The situation is very different when we approximate the dilute limit. For WDC suspensions with particle volume fraction $\phi\leq10^{-3}$ the surface potential becomes approximately constant, in contrast with the growth noticed in a pure salt-free suspension (SF). This behaviour is due to the different sources of \ce{H+} counterions: the added counterions released by the particles and those stemming from water dissociation. The first ones dominate in the high $\phi$ region, whereas the second ones do it at low particle concentration. The diminution of the surface potential in comparison with the pure salt-free case can be explained by the increase of the counterion concentration inside the cell. For WDNC suspensions (dotted lines), there is an additional decrease of the surface potential and a wider influence of water dissociation to higher volume fraction values in comparison with the case of coincident counterions, WDC (dashed lines). This is due to the fact that now the added counterions do not participate in the water dissociation reaction. Therefore, the total number of counterions is larger than in the WDC case, causing a better screening of the particle charge and, consequently, diminishing the electric potential at the particle surface.

Fig. \ref{fgr:WDC_WDNC_PHI} also shows that the inclusion of the finite ion size effect (MPB calculations) always rises the surface electric potential in comparison with the PB predictions. The reason relies on the limitation of the ionic concentration in the neighborhood of the particle, which seriously diminishes the screening of the particle charge, and consequently, leading to an increment of the surface potential. An additional significative increase of the surface potential is obtained when the distance of closest approach of the ions to the particle surface is taken into account (MPBL model). As expected, the existence of a Laplace region free of ions also penalizes the screening of the particle charge. We always find this behaviour whatever the cases studied: SF, WDC or WDNC suspensions. As we mentioned before, there is a potential shift between the red and blue lines which is independent of the volume fraction and also agrees numerically well with Equation \ref{shift}. This explain the parallelism observed between these curves.

In the MPB case without a Laplace layer, and for sufficiently highly charged particles, a region of constant charge density develops very close to the particle surface with a thickness that is approximately independent of the volume fraction. This is in contrast with the classical PB problem, where the charge density at the particle surface is unbounded and can reach unphysical values. It is clear that we can apply a similar explanation as before for the parallelism that also exists between the red and black curves.

\begin{figure}[t!]
\centering
  \includegraphics[height= 6.4cm]{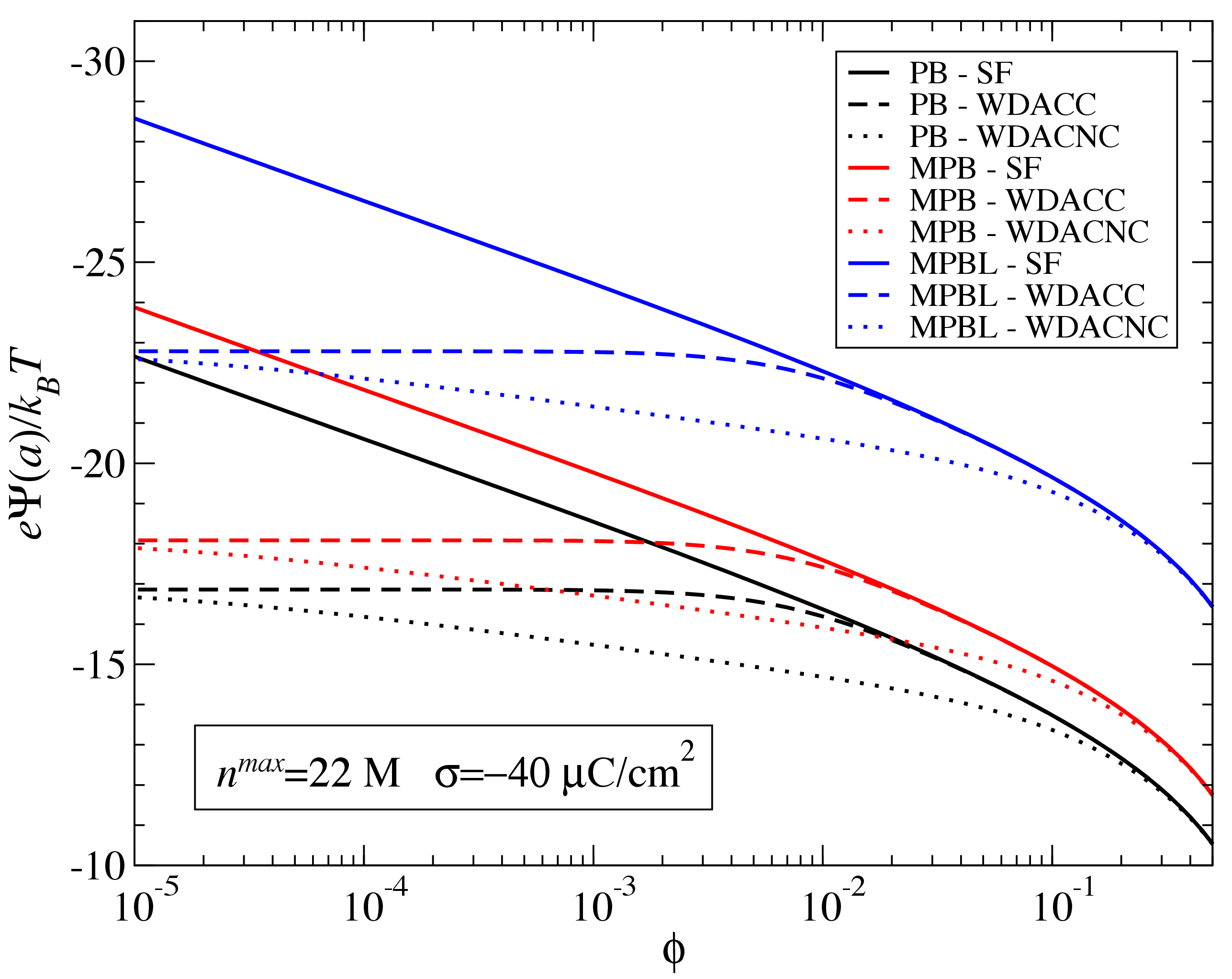}
  \caption{Dimensionless surface electric potential as a function of particle volume fraction, considering (blue lines, MPBL) or not (red lines, MPB) the excluded region in contact with the particle. Black lines show the results of the PB equation. Solid lines: only added counterions \ce{H+} (SF). Dashed lines: added counterions \ce{H+} with water dissociation and atmospheric contamination ions (WDACC). Dotted lines: added counterions different than \ce{H+} with $z_c=1$ and water dissociation and atmospheric contamination ions (WDACNC).}
  \label{fgr:WDACC_WDACNC_PHI}
\end{figure}

Fig. \ref{fgr:WDACC_WDACNC_PHI} shows the dimensionless equilibrium electric potential at the surface of the particle for a wide range of particle volume fractions. We display in black, red and blue lines the predictions of the PB, MPB and MPBL equations, respectively. Solid lines account for the results of pure salt-free suspensions with only added counterions \ce{H+} (SF). Dashed lines stand for realistic WDACC suspensions with \ce{H+} ions as added counterions, and dotted lines for realistic WDACNC suspensions with added counterions different than \ce{H+} with $z_c=1$. In the case of WDACC (dashed lines), we again observe that for highly concentrated suspensions the results coincide with those of pure salt-free predictions (SF), due to the added counterions dominance over water dissociation and atmospheric contamination ions. In contrast with that shown in Fig. \ref{fgr:WDC_WDNC_PHI} for water dissociation ions, when we also consider atmospheric \ce{CO2} contamination, the plateau in the surface potential now extends from the very dilute limit to $\phi=10^{-2}$. In the present case, the \ce{H+} counterions inside of the cell arise from three different mechanisms: the charging process of the colloidal particle, the water dissociation equilibrium, and the dissociated protons from atmospheric carbonic acid. Consequently, there is a great increase of the concentration of counterions that accounts for the marked reduction of the surface potential in the low volume fraction region.

For WDACNC suspensions (dotted lines), Fig. \ref{fgr:WDACC_WDACNC_PHI} displays an additional decrease of the surface potential and an extended influence of the atmospheric \ce{CO2} contamination for larger volume fractions in comparison with the case of WDACC suspensions. The explanation is based again on the fact that the non coincident added counterions do not participate in the water and carbonic acid dissociation reactions, yielding a large number of counterions in solution. The resulting screening of the particle charge is enhanced in comparison with the WDACC case, and therefore, the surface potential decreases.

Regarding the finite ion size effect, Fig. \ref{fgr:WDACC_WDACNC_PHI} shows that the surface potential always increases for MPB calculations in comparison with the PB predictions. This is again due to the limitation of the ionic concentration in the neighborhood of the particle surface, which provokes a diminution of the screening of the particle charge, and consequently, raising the surface potential. The surface potential increases in a significative way when we consider a distance of closest approach of the ions to the particle surface (MPBL model). The reason lies on the reduction of the particle charge screening as a consequence of the Laplace region free of ions next to the surface. We again find this behaviour irrespective of the cases studied: SF, WDACC or WDACNC suspensions.

\begin{figure}[t!]
\centering
  \includegraphics[height= 6.4cm]{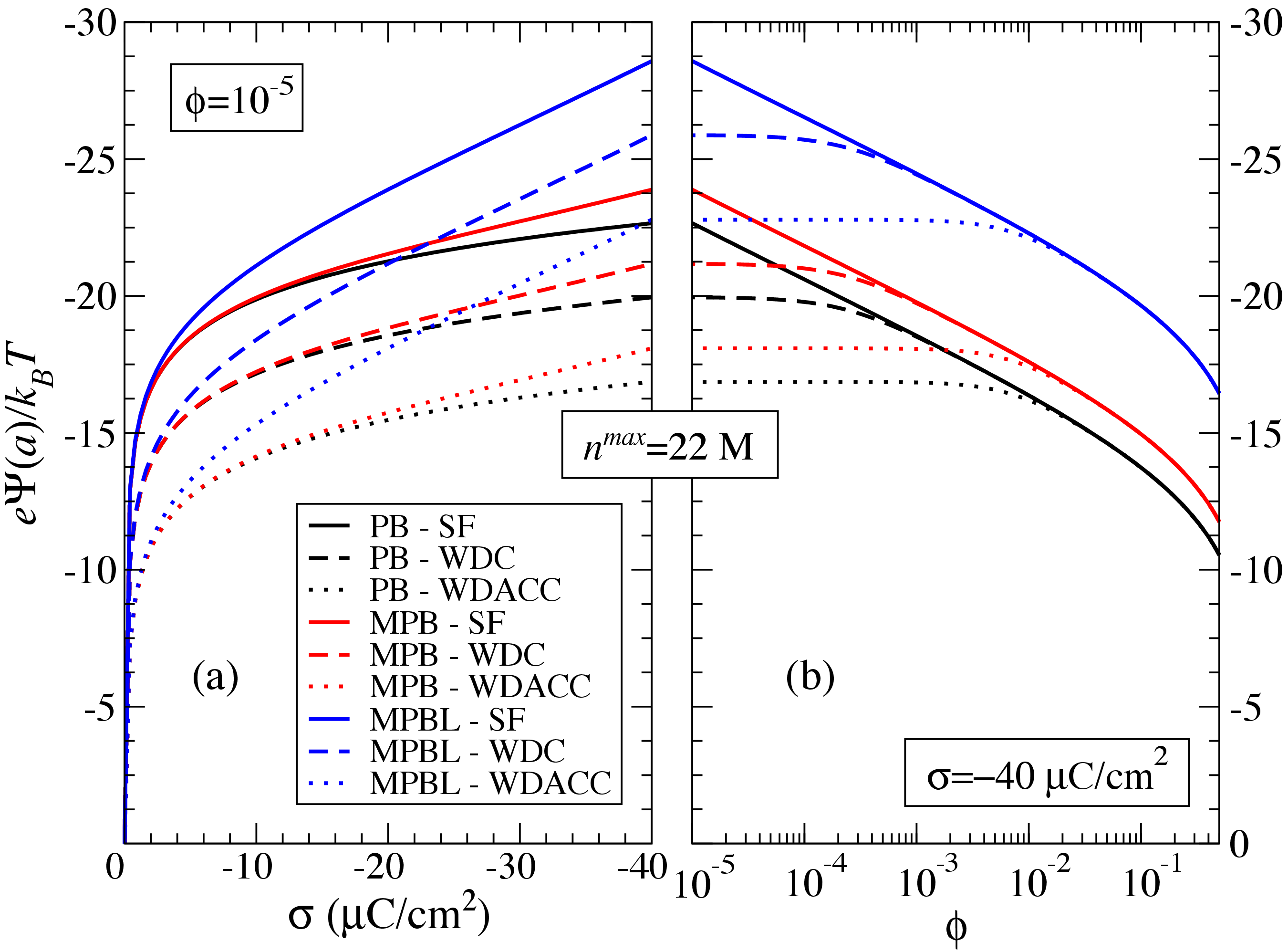}
  \caption{Dimensionless surface electric potential for different values of particle surface charge density (a) and particle volume fraction (b), considering (blue lines, MPBL) or not (red lines, MPB) the excluded region in contact with the particle. Black lines show the results of the PB equation. Solid lines: only added counterions \ce{H+} (SF). Dashed lines: added counterions \ce{H+} with water dissociation ions (WDC). Dotted lines: added counterions \ce{H+} with water dissociation and atmospheric contamination ions (WDACC).}
  \label{fgr:WDC_WDACC_SIGPHI}
\end{figure}

Fig. \ref{fgr:WDC_WDACC_SIGPHI} displays the dimensionless equilibrium surface electric potential for a wide range of particle surface charge densities, Fig. \ref{fgr:WDC_WDACC_SIGPHI}a, and particle volume fractions, Fig. \ref{fgr:WDC_WDACC_SIGPHI}b. We repeat this study for different salt-free suspensions, SF, WDC and WDACC, with added counterions coincident with \ce{H+}, and for the three PB, MPB and MPBL models. We take the maximum possible ionic concentration due to the excluded volume effect as $n^{max}=22$ M, corresponding to a hydrated hydronium ion in solution. We observe in Fig. \ref{fgr:WDC_WDACC_SIGPHI}a that the diminution of the surface potential due to realistic considerations (dashed or dotted lines against solid lines) is practically independent of the particle surface charge density, for both point-like and finite size ions, even when considering the excluded region in contact with the particle. The results of Fig. \ref{fgr:WDC_WDACC_SIGPHI}b confirm those from Figs. \ref{fgr:WDC_WDNC_PHI} and \ref{fgr:WDACC_WDACNC_PHI}, and we can clearly see how the plateaus in the surface potential extend to larger volume fractions when we consider WDACC (dotted lines) instead of WDC (dashed lines) suspensions.

\begin{figure}[t!]
\centering
  \includegraphics[height= 6.4cm]{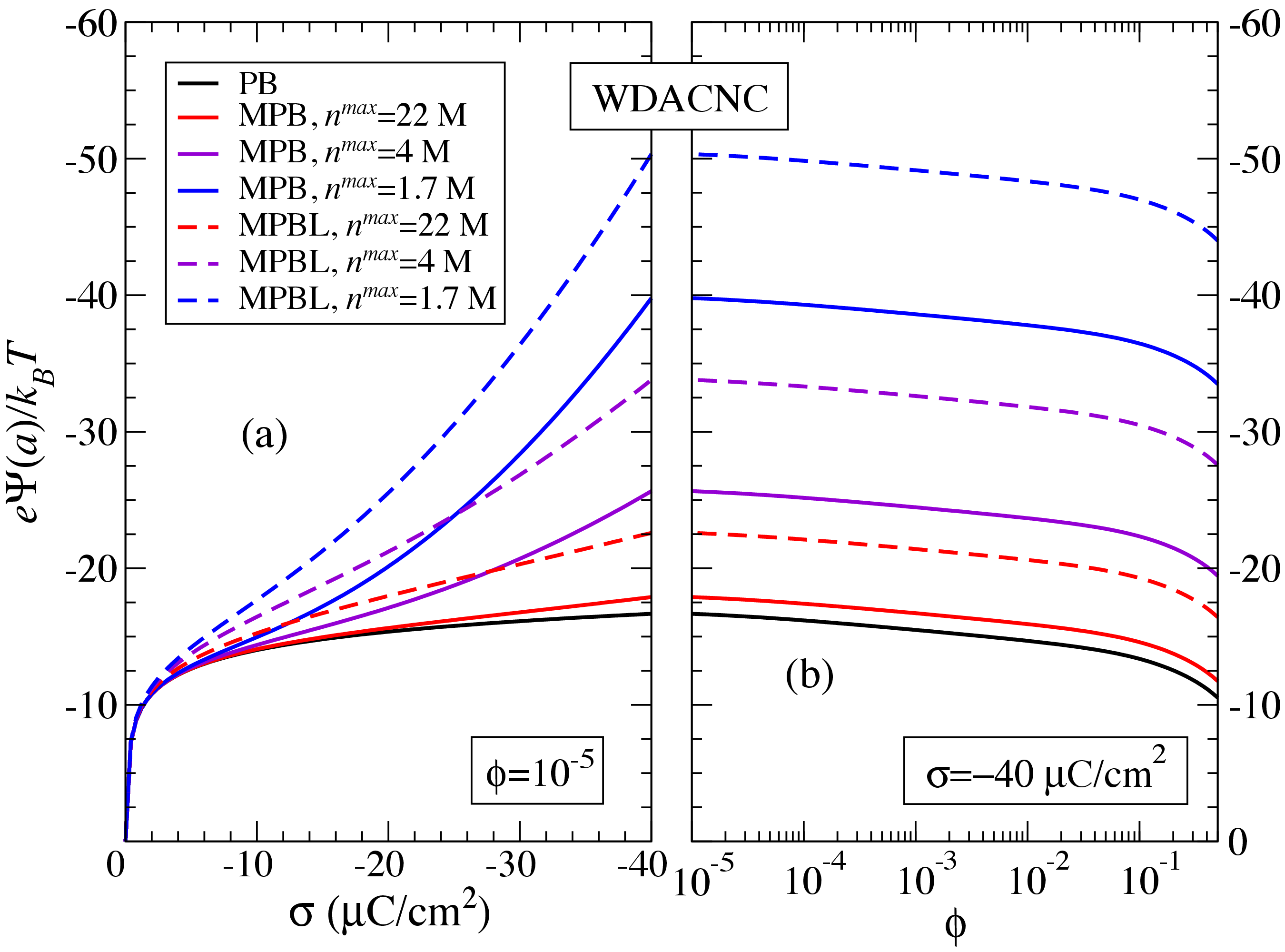}
  \caption{Dimensionless surface electric potential for different values of particle surface charge density (a) and particle volume fraction (b) for different ion sizes, considering (dashed lines, MPBL) or not (solid lines, MPB) the excluded region in contact with the particle. Black lines show the results of the PB equation. In all cases we consider salt-free suspensions with added counterions different than \ce{H+} with $z_c=1$ and water dissociation and atmospheric contamination ions (WDACNC).}
  \label{fgr:WDACNC_NMAX_SIGPHI}
\end{figure}

Fig. \ref{fgr:WDACNC_NMAX_SIGPHI} displays the dimensionless equilibrium surface electric potential for a wide range of particle surface charge densities, Fig. \ref{fgr:WDACNC_NMAX_SIGPHI}a, and particle volume fractions, Fig. \ref{fgr:WDACNC_NMAX_SIGPHI}b. We repeat this study for different ion sizes. In all cases we examine realistic salt-free suspensions with added counterions different than \ce{H+} with $z_c=1$ and water dissociation and atmospheric contamination ions (WDACNC). We find that the surface electric potential increases with the particle charge density, Fig. \ref{fgr:WDACNC_NMAX_SIGPHI}a. However, in some cases there is a different behaviour in comparison with the point-like case. Initially, a fast and roughly increase of the surface potential with the surface charge density is observed, which is followed by a much slower growth at higher surface charge densities for the PB case, or when the size of the ions is very small. This phenomenon is related to the classical counterion condensation effect: for high surface charges a layer of counterions develops very close to the particle surface \cite{Ohshima2006}. When the ion size is taken into account (MPB), we limit the appearance of the classical condensation effect because when the surface charge is increased, the additional counterions join the condensate enlarging it. If a region of closest approach of the ions to the particle surface is also considered (MPBL), the mechanism is the same, but now the additional counterions are located in farther positions from the particle surface. This explains the additional increase of the surface potential observed for large ion sizes and high surface charges densities. These profiles are similar to those found by the authors for pure salt-free suspensions \cite{Roa2011}, but the values of the surface potential are lower for WDACNC suspensions than for SF suspensions, as Fig. \ref{fgr:WDACC_WDACNC_PHI} displayed.

On the other hand, the surface electric potential decreases when the particle volume fraction increases, irrespective of the cases studied: PB, MPB or MPBL, Fig. \ref{fgr:WDACNC_NMAX_SIGPHI}b. When the particle concentration raises, the available space for the ions inside the cell decreases and, consequently, the screening of the particle charge largely augments, thus reducing the value of the surface potential.

\begin{figure}[t!]
\centering
  \includegraphics[height= 6.4cm]{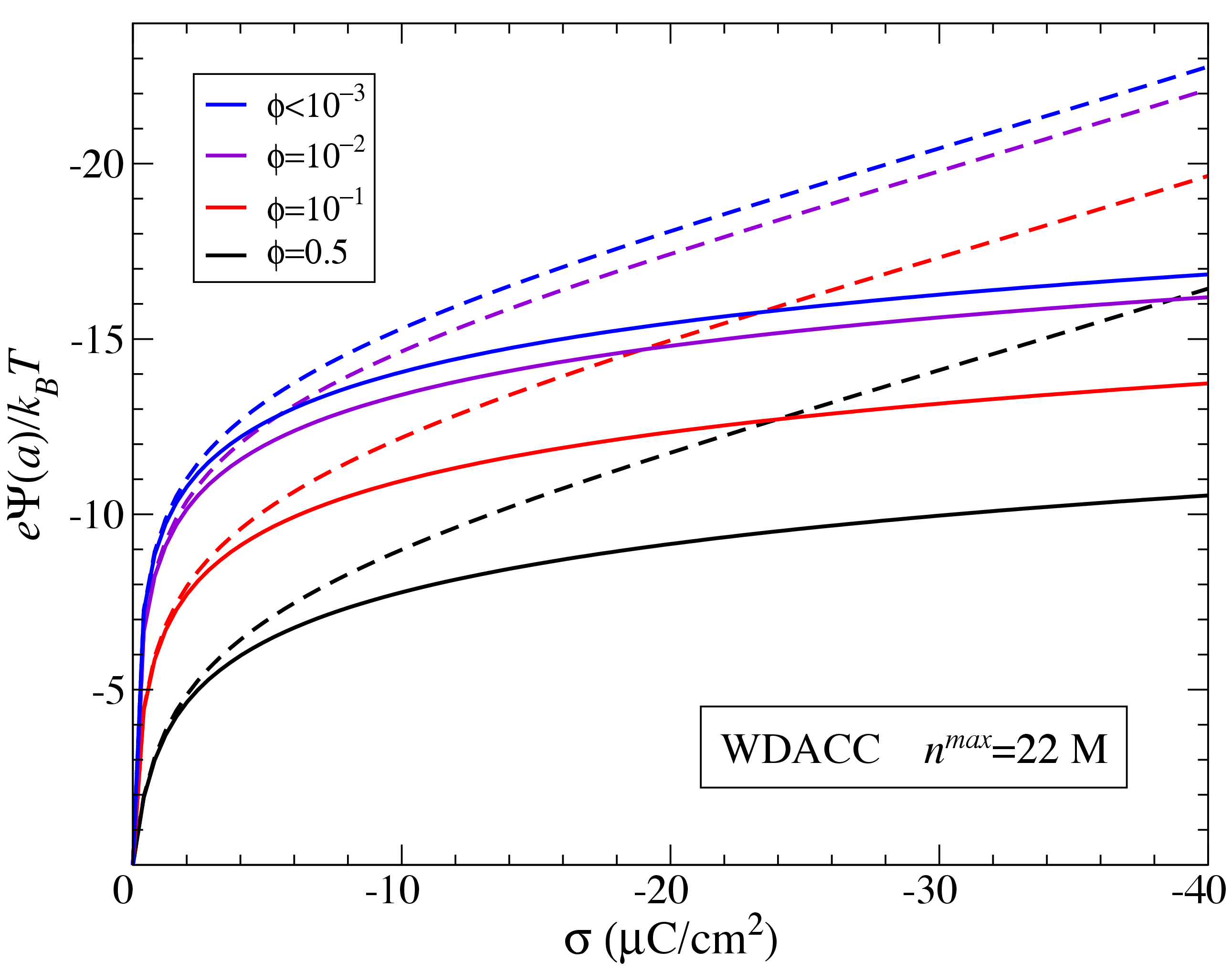}
  \caption{Dimensionless surface potential against the surface charge density for different particle volume fraction values. Solid lines stand for the results of the PB equation. Dashed lines show the results of the MPBL model. In all cases we consider salt-free suspensions with added counterions \ce{H+} and water dissociation and atmospheric contamination ions (WDACC).}
  \label{fgr:WDACC_DPHI_SIG}
\end{figure}

\begin{figure}[t!]
\centering
  \includegraphics[height= 6.4cm]{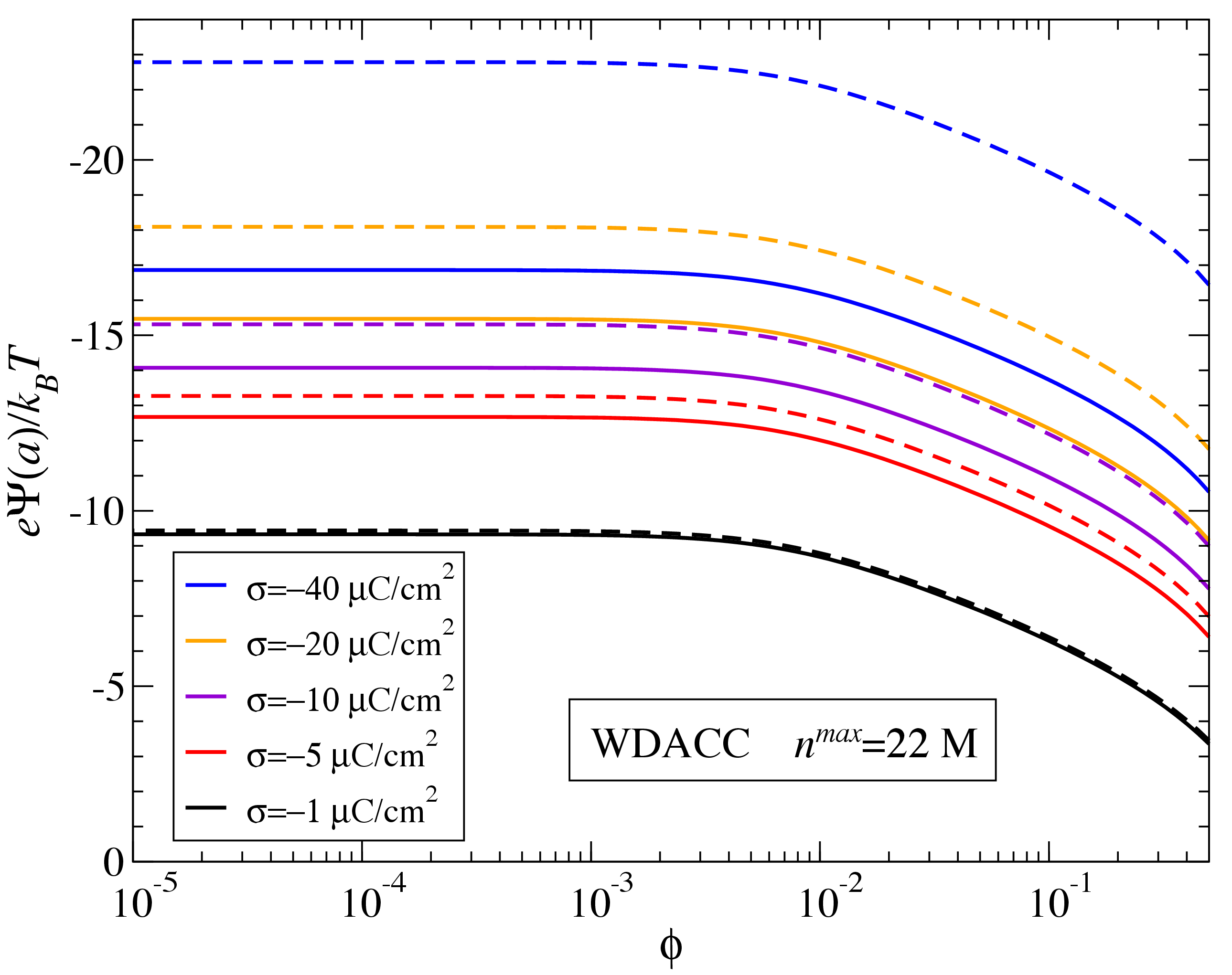}
  \caption{Dimensionless surface potential against the particle volume fraction for different surface charge density values. Solid lines stand for the results of the PB equation. Dashed lines show the results of the MPBL model. In all cases we consider salt-free suspensions with added counterions \ce{H+} and water dissociation and atmospheric contamination ions (WDACC).}
  \label{fgr:WDACC_DSIG_PHI}
\end{figure}

Figs. \ref{fgr:WDACC_DPHI_SIG} and \ref{fgr:WDACC_DSIG_PHI} expand the results of Fig. \ref{fgr:WDC_WDACC_SIGPHI} for realistic salt-free suspensions with added counterions coincident with \ce{H+} and water dissociation and atmospheric contamination ions (WDACC). In them, we compare the results of the PB equation (solid lines) with those of the MPBL model (dashed lines) for a given ion size (typical of an hydronium ion in solution). Fig. \ref{fgr:WDACC_DPHI_SIG} presents the dimensionless equilibrium surface electric potential at a wide range of particle surface charge densities. The different coloured lines correspond to different particle volume fraction. The most remarkable fact shown in Fig. \ref{fgr:WDACC_DPHI_SIG} is the large influence of the finite ion size effect even for moderately low particle surface charge densities. While for PB predictions (solid lines) the surface potential hardly increases with surface charge for moderate to high surface charges at each volume fraction, the MPBL results (dashed lines) display an outstanding growth for the same conditions, associated with the lower charge screening ability of the counterions because of their finite size. This fact will surely have important consequences on the electrokinetic properties of such particles in concentrated realistic salt-free suspensions, as has already been shown for dilute suspensions in electrolyte solutions by Aranda-Rasc\'on \textit{et al.} \cite{ArandaRascon2009,ArandaRascon2009b}. 

Finally, Fig. \ref{fgr:WDACC_DSIG_PHI} shows the dimensionless equilibrium electric potential at the particle surface against the particle volume fraction. The different coloured lines correspond to different negative particle surface charge densities. As previously stated, the particle surface potential shows an initial plateau in the dilute region followed by a monotonous decrease with volume fraction at fixed particle charge density. For the ionic size chosen, the larger the surface charge, the larger the relative increase of the surface potential at every volume fraction, due to the enlarging of the counterion condensate in the neighborhood of the particle surface.

From the results, we think that it is clear that the influence of the finite ion size effect on the EDL description cannot be neglected for many typical particle charges and volume fractions, either in a pure or in a realistic salt-free suspension.

\section{Conclusions}

In this work we have studied the influence of finite ion size corrections on the description of the equilibrium electric double layer of a spherical particle in a realistic salt-free concentrated suspension, in an attempt to get us closer to real systems. These realistic suspensions include water dissociation ions and those generated by atmospheric carbon dioxide contamination, in addition to the added counterions released by the particles to the solution. The resulting model is based on a mean-field approach that has reasonable succeeded in modeling electrokinetic and rheological properties of concentrated suspensions.

We have used a cell model approach to account for particle-particle interactions, and derived modified Poisson-Boltzmann equations (MPB and MPBL) which include such ion size effects. The theoretical procedure has followed that by Borukhov\cite{Borukhov2004} with the additional inclusion of an excluded region of closest approach of the ions to the particle surface \cite{ArandaRascon2009,ArandaRascon2009b}.  The results have shown that the finite ion size effect (MPB equation) has to be taken into account for moderate to high particle charges at every particle volume fraction, and even more if a distance of closest approach of the ions to the particle surface is considered (MPBL model), irrespective of the realistic models (WDC, WDNC, WDACC or WDACNC) used.

The equilibrium model presented in this paper will be used to develop nonequilibrium models of the response of a realistic salt-free concentrated suspension to external electric fields. Experimental results concerning the DC electrophoretic mobility, dynamic electrophoretic mobility, electrical conductivity and dielectric response, should be compared with the predictions of these future models to test them. To carry out such comparisons highly charged particles like those of some sulfonated polystyrene latexes might be used. These theoretical and experimental tasks will be addressed by the authors in the near future.

\section*{Acknowledgements}
Junta de Andaluc\'ia, Spain (Project P08-FQM-3779), MEC, Spain (Project FIS2007-62737) and MICINN, Spain (Project FIS2010-18972), co-financed with FEDER funds by the EU. Helpful discussions with Dr. Juan J. Alonso are also gratefully acknowledged.

\end{document}